\documentclass[11pt,aps,prd,superscriptaddress,reprint,twocolumn,preprintnumbers,nofootinbib,floatfix]{revtex4-2}
\usepackage{microtype}
\usepackage{booktabs,graphicx,xcolor,bm}
\usepackage[colorlinks=true,linkcolor=blue!60!black,citecolor=blue!60!black,urlcolor=blue!60!black]{hyperref}

\usepackage[utf8]{inputenc}
\usepackage{bbold}
\usepackage{comment}
\usepackage[normalem]{ulem}  
\usepackage{xcolor}
\usepackage{comment}

\usepackage{subcaption}  
\usepackage{afterpage}

\usepackage{amsmath,amssymb,amsfonts}

\usepackage{physics}
\usepackage{hyperref}
\usepackage[capitalise]{cleveref}

\definecolor{mycolor}{RGB}{90, 160, 255}
\hypersetup{
    colorlinks=true,       
    linkcolor=mycolor,        
    citecolor=mycolor,       
    urlcolor=cyan          
}

\newcommand{\dmsqSL}{\Delta m_{21}^2}
\newcommand{\dmsqATM}{\Delta m_{31}^2}
\newcommand{\snqS}{\sin^2\theta_{12}}
\newcommand{\snqR}{\sin^2\theta_{13}}
\newcommand{\unit}{10^{-3}\,\text{eV}^2}
\newcommand{\TMO}{T_{\rm MO}}
\newcommand{\chisq}{\chi^2}
\newcommand{\dchisq}{\Delta\chi^2}
\newcommand{\dmsqmm}{\Delta m_{\mu\mu}^2}
\newcommand{\dmsqee}{\Delta m_{ee}^2}
\newcommand{\orcid}[1]{\href{https://orcid.org/#1}{#1}}
\newcommand{\AccIC}{Acc+IC}

\begin{document}

\preprint{FERMILAB-PUB-26-0812-T}

\title{Where does the neutrino Mass Ordering preference come from?}

\author{Pedro A.~N.~Machado}
\email{pmachado@fnal.gov}
\thanks{\orcid{0000-0002-9118-7354}}
\affiliation{Theory Division, Fermi National Accelerator Laboratory, Illinois 60510, USA}

\author{João Paulo Pinheiro}
\email{joaopaulo.pinheiro@sjtu.edu.cn}
\thanks{\orcid{0000-0002-6536-2040}}
\affiliation{State Key Laboratory of Dark Matter Physics, Tsung-Dao Lee Institute \& School of Physics and Astronomy, Shanghai Jiao Tong University, Shanghai 200240, China}
\affiliation{Key Laboratory for Particle Astrophysics and Cosmology (MOE) \& Shanghai Key Laboratory for Particle Physics and Cosmology, Shanghai Jiao Tong University, Shanghai 200240, China}
\date{\today}

\begin{abstract}
After the recent JUNO result, the comparison of the effective mass splittings measured in $\bar\nu_e$ disappearance at reactors and in $\nu_\mu$ disappearance has become one of the leading handles on the neutrino mass ordering.
In this paper, we study which data sets drive the present preference for the ordering, combining JUNO and Daya Bay with $\nu_\mu$ disappearance data from NOvA, T2K, MINOS, and IceCube/DeepCore.
While the combination disfavors the inverted ordering at $1.9\sigma$, this preference comes from a balance between several small preferences. 
In particular, the combinations of reactor data with T2K or with NOvA point to opposite orderings.
The values of the mass splitting preferred by these two experiments differ by about $3.5\%$, comparable to the difference expected between the two orderings, and thus a shift of this size in a single experiment is enough to reverse the preference.
Moreover, in the context of this combination of reactor and $\nu_\mu$ disappearance data, further JUNO data would not bring the rejection of the inverted ordering beyond about $3\sigma$, as the comparison is then limited by the precision of the $\nu_\mu$ disappearance experiment.
Improving the $\nu_\mu$ disappearance measurements requires control of the neutrino energy reconstruction, and thus the modeling of neutrino-nucleus interaction physics, at the percent level.
\end{abstract}
\maketitle

\section{Introduction}
\label{sec:intro}

\begin{figure*}[t!]
\centering
\includegraphics[width=\linewidth]{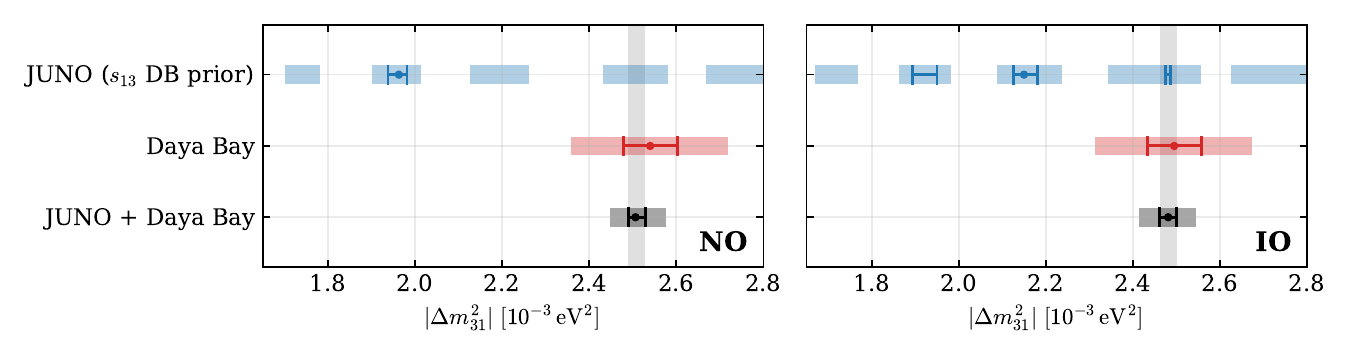}
\caption{Allowed values of $|\dmsqATM|$ in NO (left) and IO (right) from JUNO (blue), Daya Bay (red) and their combination (black).
Error bars show the $1\sigma$ intervals and shaded bands the $3\sigma$ intervals; the vertical gray band is the $1\sigma$ interval of JUNO combined with Daya Bay.
The intervals of JUNO and of its combination with Daya Bay are obtained with the Feldman--Cousins construction.
For JUNO, $\snqR$ is marginalized over with a Gaussian prior from Daya Bay.
}
\label{fig:dm31_reactor}
\end{figure*}   

Neutrino oscillation experiments have established that neutrinos are massive and that lepton flavors mix, and most of the oscillation parameters are now known at the few-percent level~\cite{Super-Kamiokande:1998kpq, Super-Kamiokande:2023ahc, IceCubeCollaboration:2023wtb, SNO:2002tuh, SNO:2011hxd, Super-Kamiokande:2023jbt, Borexino:2017rsf, KamLAND:2002uet, KamLAND:2010fvi, DayaBay:2022orm, RENO:2018dro, JUNO:2025gmd, JUNO:2026tap, MINOS:2020llm, T2K:2025yoy, NOvA:2025tmb}.
Nevertheless, three open questions stand out within the three-flavor framework, namely the value of the $CP$-violating phase $\delta_{\rm CP}$, the octant of $\theta_{23}$, and the neutrino mass ordering, that is, whether $m_3$ is the heaviest (normal ordering, NO) or the lightest (inverted ordering, IO) of the three masses.
The mass ordering in particular is of major importance to neutrino physics and to physics beyond oscillations.
In cosmology, the ordering sets the minimum of the sum of neutrino masses, about $60$~meV in NO and $100$~meV in IO.
Interestingly, the bound from baryon acoustic oscillations combined with cosmic microwave background data already lies below the IO minimum, $\sum m_\nu<64$~meV at $95\%$~CL~\cite{Planck:2018vyg,DESI:2025zgx,Elbers:2025vlz}.
In neutrinoless double beta decay, IO implies an effective Majorana mass above about $20$~meV, a region that the next generation of experiments will explore, while in NO this mass could vanish if the Majorana phases conspire to that end~\cite{Agostini:2022zub,KamLAND-Zen:2024eml,LEGEND:2025jwu}.
Similarly, the ordering sets the minimum of the effective mass probed in kinematic measurements,  $50$~meV for IO versus $9$~meV for NO, both below the present KATRIN limit of $0.45$~eV~\cite{KATRIN:2024cdt}.

Oscillation experiments can determine the mass ordering through matter effects, through the interference of the two fast oscillation frequencies, or through the comparison of effective mass splittings measured in different channels.
Matter effects enhance $\nu_\mu\to\nu_e$ oscillations for neutrinos in NO and for antineutrinos in IO, and thus are best probed by experiments in which neutrinos cross a large amount of matter.
This is realized in atmospheric neutrino samples in Super-Kamiokande, IceCube and KM3NeT ~\cite{Super-Kamiokande:2023ahc,IceCubeCollaboration:2023wtb,KM3NeT:2024ecf}, as well as in the future DUNE experiment~\cite{DUNE:2020ypp,DUNE:2020jqi}.
In reactor antineutrinos at a baseline of about $50$~km, the fast oscillations driven by $\dmsqATM$ and $\Delta m^2_{32}$ interfere, and the ordering decides which of the two frequencies carries the larger amplitude~\cite{Petcov:2001sy,Choubey:2003qx,Learned:2006wy,Li:2013zyd}.
This is the approach of JUNO~\cite{JUNO:2015zny,JUNO:2024jaw}; it is a precision measurement, as the effect is a small shift in the phase of the fast oscillations, and it requires an energy resolution at the percent level and an even more precise control of the energy scale.
Lastly, one may compare the effective mass splittings measured in $\bar\nu_e$ disappearance at reactors, $\dmsqee$, and in $\nu_\mu$ disappearance at accelerator and atmospheric experiments, $\dmsqmm$~\cite{deGouvea:2005hk,Nunokawa:2005nx,Minakata:2006gq,Parke:2024xre}.
For normal ordering $|\dmsqee|>|\dmsqmm|$, while the opposite is true for inverted ordering. 
The difference between the two effective splittings is of few percent, depending on mixing angles and the $CP$ phase.
This last approach has gained attention after the first JUNO data, being used by the JUNO collaboration~\cite{JUNO:2026tap}, by NOvA~\cite{NOvA:2026tir}, and in global analyses~\cite{Esteban:2026phq}, and it is the focus of this paper.

This way of determining the ordering requires precise measurements of both splittings, as single effective splitting carries no information on the ordering.
While JUNO combined with Daya Bay already measures $|\dmsqee|$ with an uncertainty below $1\%$, each accelerator experiment measures $|\dmsqmm|$ with an uncertainty of about $2\%$ so that each one can provide at most a mild preference for the ordering.
As a consequence, the preferred ordering depends on the value of the splitting that each experiment extracts from its data, and a shift of about one percent in a single experiment, from a statistical fluctuation or from the modeling of its energy scale, can change the outcome.

Since the comparison of $\dmsqee$ and $\dmsqmm$ is now one of the leading handles on the mass ordering, it is important to understand which data sets drive the present preference and how robust it is.
In this paper, we analyze this comparison using the reactor experiments JUNO and Daya Bay and the accelerator experiments T2K and NOvA, together with MINOS, IceCube/DeepCore and their combination in NuFIT~6.1~\cite{Esteban:2024eli,Esteban:2026phq}.
Following each experiment separately, we identify where the data agree, where they disagree, and which of them push the preference towards each ordering.
In Sec.~\ref{sec:measurements}, we discuss how the mass splittings are extracted from reactor and accelerator data, and in Sec.~\ref{sec:combination} we combine them and quantify the preference for each ordering.
We then discuss the impact of future JUNO data in Sec.~\ref{sec:future}, and conclude in Sec.~\ref{sec:conclusion}.
Since the likelihood of JUNO has several nearly degenerate minima in $\dmsqATM$, the usual asymptotic formulas do not apply, and all significances are obtained from pseudo-experiments; the statistical treatment is described in Appendix~\ref{app:stat}.

\section{Measurements of $\dmsqee$ and $\dmsqmm$}
\label{sec:measurements}

\begin{figure*}[t!]
\centering
\includegraphics[width=\linewidth]{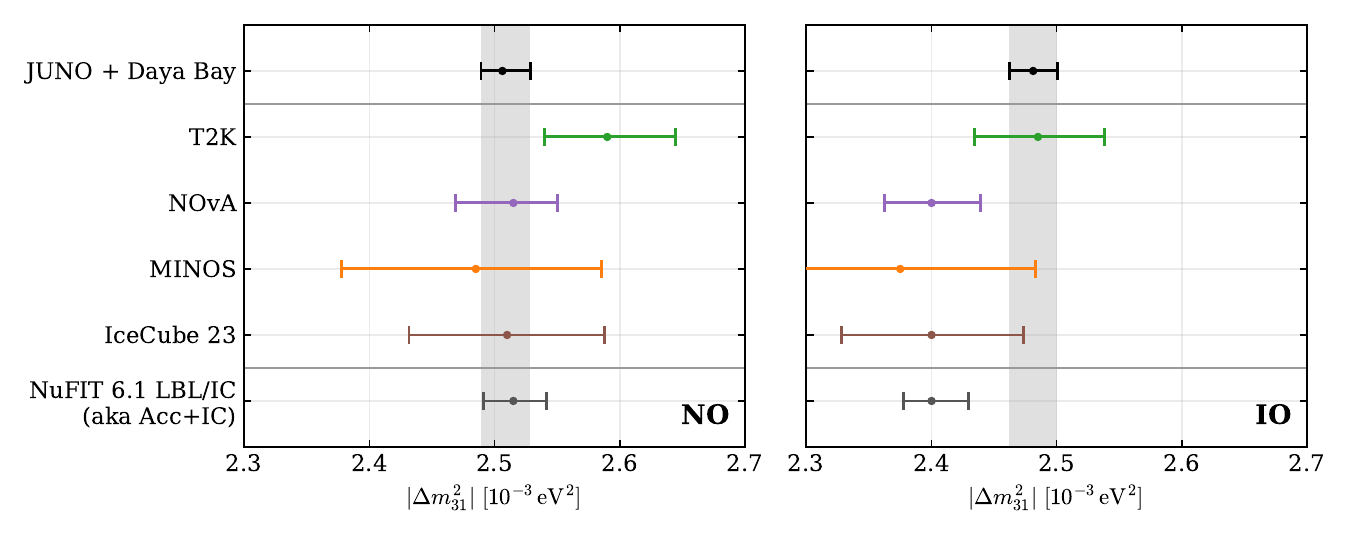}
\caption{$1\sigma$ allowed values of $|\dmsqATM|$ in NO (left) and IO (right) from JUNO combined with Daya Bay (black), T2K (green), NOvA (purple), MINOS (orange), IceCube/DeepCore (brown) and their NuFIT~6.1 combination, \AccIC{} (gray).
The vertical gray band is the $1\sigma$ interval of JUNO combined with Daya Bay.
}
\label{fig:dm31_comparison}
\end{figure*}

When the baseline and energy of a disappearance experiment are chosen near the first oscillation maximum driven by $\dmsqATM$, as in Daya Bay, T2K and NOvA, the survival probability is well described by a single effective mass splitting, which depends on the flavor that disappears.
For $\bar\nu_e$ disappearance at reactors and $\nu_\mu$ disappearance at accelerators, these effective splittings read~\cite{Nunokawa:2005nx}
\begin{align}
\dmsqee &\simeq \cos^2\theta_{12}\,\dmsqATM+\sin^2\theta_{12}\,\Delta m^2_{32}, \label{eq:dmee}\\
\dmsqmm &\simeq \sin^2\theta_{12}\,\dmsqATM+\cos^2\theta_{12}\,\Delta m^2_{32}\\
&\quad+\cos\delta_{\rm CP}\sin\theta_{13}\sin2\theta_{12}\tan\theta_{23}\,\dmsqSL, \label{eq:dmmm}
\end{align}
and their difference is
\begin{align}
&|\dmsqee| - |\dmsqmm| = \label{eq:dmee_dmmm}\\
&\quad\pm \dmsqSL
\left(\cos2\theta_{12}-\cos\delta_{\rm CP}\sin\theta_{13}\sin2\theta_{12}\tan\theta_{23}\right),\nonumber
\end{align}
with the upper sign for NO and lower for IO.
Nevertheless, experiments report $\dmsqATM$ or $\Delta m^2_{32}$, fitted under an assumption on the ordering, and we will follow this convention and present all results in terms of $|\dmsqATM|$.

We begin with JUNO,
which detects reactor antineutrinos at a baseline of $52.5$~km, near the first solar oscillation maximum.
The slow oscillation of its spectrum determines $\dmsqSL$ and $\theta_{12}$~\cite{JUNO:2025gmd,JUNO:2026tap}.
Its energy resolution also allows it to observe the fast oscillations driven by $\dmsqATM$, which appear as small wiggles on top of the slow one.
The frequency of these wiggles, however, is not determined unambiguously.
A shift in $\dmsqATM$ changes the oscillation phase $\dmsqATM L/4E$, and when this change is a multiple of $\pi$ at the energies where JUNO is most sensitive, the wiggles in the spectrum are almost unchanged.
With the present statistics, JUNO cannot exclude these approximate degeneracies, which appear as a series of minima of its $\chisq$ in $\dmsqATM$, separated by about $0.2\times\unit$.

Daya Bay, on the other hand, measured reactor antineutrinos at baselines up to about $1.6$~km, near the first oscillation maximum driven by $\dmsqATM$, and thus is free from this ambiguity~\cite{DayaBay:2022orm}.
Its measurement of $|\dmsqATM|$ selects the JUNO minimum close to $2.5\times\unit$, and the combination of the two experiments determines $|\dmsqee|$ with an uncertainty below $1\%$.
We show in \cref{fig:dm31_reactor} the 1$\sigma$ and 3$\sigma$ allowed regions for $|\dmsqATM|$ for JUNO and Daya Bay, as well as their combination.
Because of the multiple minima of the JUNO $\chisq$, the conditions for Wilks' theorem are strongly violated, and we obtain the intervals of JUNO with the Feldman--Cousins construction~\cite{Feldman:1997qc}, in which the $\dchisq$ cut is computed point by point from pseudo-experiments.
For JUNO, following the procedure of the collaboration, we marginalize over $\sin^2\theta_{13}$ with a Gaussian prior from Daya Bay.

We can see that while locally JUNO has smaller error bars, there are several degeneracies at the 3$\sigma$ level, and there is a slight disagreement between these experiments due to the JUNO preference for a low value of $|\dmsqATM|$.
For the same reason, we compute the significance of this tension with pseudo-experiments, finding $1.4\sigma$ in NO and $1.0\sigma$ in IO (see Appendix~\ref{app:stat} for details on simulations and statistical treatment).
This tension is comparable for both orderings.
Since JUNO and Daya Bay measure the same effective splitting, the reactor data alone show no preference for either ordering.

The most precise measurements of $\nu_\mu$ disappearance come from T2K and NOvA, at baselines of $295$ and $810$~km, 
respectively, with beams peaked near the first oscillation maximum driven 
by $\dmsqATM$~\cite{T2K:2025yoy,NOvA:2025tmb}.
MINOS~\cite{MINOS:2020llm} and IceCube/DeepCore~\cite{IceCubeCollaboration:2023wtb} 
also measure this splitting with muon neutrinos.
For each of these experiments, we use the one-dimensional $\chisq$ profile in 
$\dmsqATM$ from NuFIT~6.1~\cite{Esteban:2024eli,Esteban:2026phq}, obtained from 
the full data of the experiment, appearance and disappearance samples together, 
with $\snqR=0.0222$ and the solar parameters fixed and $\theta_{23}$ and 
$\delta_{\rm CP}$ marginalized over.
We also use the NuFIT~6.1 combination of MINOS, T2K, NOvA and the 9.3-year 
IceCube/DeepCore sample, which we denote Acc+IC.
Atmospheric data from Super-Kamiokande are not included in this combination.
In \cref{fig:dm31_comparison} we compare the allowed ranges of 
$|\dmsqATM|$ from all these data sets with the reactor one, under each mass ordering.

In NO, left panel of \cref{fig:dm31_comparison}, it is interesting to note that the NOvA measurement agrees with the reactor value, while T2K's lies above it.
In IO, right panel, the situation is reversed: T2K agrees with the reactor value, while NOvA lies below it.
MINOS and IceCube/DeepCore have larger uncertainties and overlap with the reactor value in both orderings.
Therefore, the two most precise $\nu_\mu$ disappearance experiments, when combined with reactor measurements, point to different orderings.

\section{Mass Ordering from the Combined Data}
\label{sec:combination}

We combine reactor and accelerator data under each ordering, 
${\rm O}={\rm NO}$ or IO, into
\begin{equation}
\chisq_{\rm O}(x)=\chisq_{J}({\rm O},x)+\sum_i G_i^{\rm O}(x),
\label{eq:chisq_comb}
\end{equation}
where $x$ denotes the oscillation parameters, $\chisq_J$ is the $\chisq$ of the JUNO spectrum, 
which we compute with our own simulation, and $G_i^{\rm O}$ is the published, almost gaussian, $\dchisq$ profile of experiment $i$, 
namely the two-dimensional profile of Daya Bay in $(\dmsqATM,\snqR)$ and the one-dimensional profiles 
of the $\nu_\mu$ disappearance experiments in $\dmsqATM$.
The latter are shifted to have zero minimum in each ordering.
To a good approximation, this removes the hint for the mass ordering coming from the impact of matter effects on the appearance channels.

We then compare the two orderings through
\begin{equation}
\TMO=\min_x\chisq_{\rm IO}(x)-\min_x\chisq_{\rm NO}(x),
\label{eq:tmo}
\end{equation}
which is positive when NO describes the data better.
Since the $\chisq$ of JUNO has several nearly degenerate minima, 
we obtain the distribution of $\TMO$ from pseudo-experiments generated 
under each ordering, and we quote the preference with the $CL_s$ 
method~\cite{Read:2002hq,Blennow:2013oma}.
In this approach, the
$p$-value of the tested ordering is normalized by the probability of
the observed outcome under the alternative ordering, and thus the rejection of one
ordering depends on how well the data agree with the other (see Appendix~\ref{app:stat}).

To understand what $\TMO$ measures, it is useful to follow 
how the minimum of $\chisq_{\rm O}$ is built.
In each ordering, this minimum is the sum of the minima of the 
individual experiments plus the tension between them, that is, 
the increase in $\chisq$ needed to make them agree on the same value 
of $|\dmsqATM|$. 
JUNO alone prefers NO by $1.8$ units of $\chisq$ (see Appendix~\ref{app:stat}).
Nevertheless, the combination of JUNO with Daya Bay leads to a tension from moving JUNO from its global minimum, at
$1.96\times\unit$ in NO and $2.15\times\unit$ in IO, to the minimum near
$2.5\times\unit$ selected by Daya Bay.
That tension amounts to $5.6$ and $4.0$ units of $\chisq$ in NO and IO, which almost exactly compensates the JUNO preference.
Overall, JUNO and Daya Bay together yield $\TMO=0.21$ and therefore no preference for either ordering.

\begin{figure}[t]
\centering
\includegraphics[width=\linewidth]{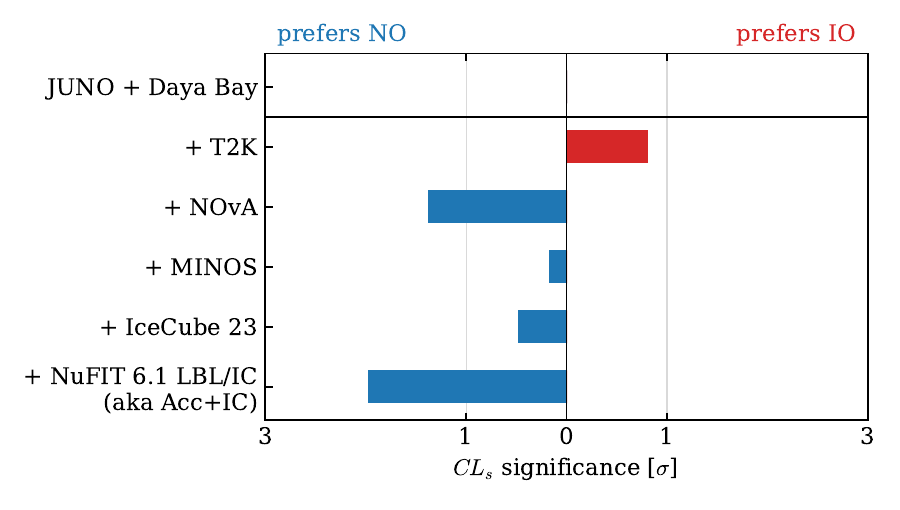}
\caption{Preference for NO (blue, left) or IO (red, right) from JUNO combined with Daya Bay, alone and together with each $\nu_\mu$ disappearance data set, expressed as the one-sided significance of $CL_s$ obtained from pseudo-experiments.}
\label{fig:mo_preference}
\end{figure}

One could also repeat this exercise for each measurement of the mass splitting in the $\nu_\mu$ disappearance channels.
We present these results in Fig.~\ref{fig:mo_preference},
in which the bars to the left (blue) indicate a preference for NO 
and the bars to the right (red) a preference for IO, for the combination of JUNO and Daya Bay plus each experiment individually and \AccIC.
Combining reactor and $\nu_\mu$ disappearance data disfavors IO at $1.9\sigma$.
We find good agreement with the official result of the JUNO 
collaboration~\cite{JUNO:2026tap} wherever a direct comparison is possible,%
\footnote{The JUNO collaboration reports $\TMO=4.9$ and a rejection of IO at $2.1\sigma$ with the one-sided $p$-value, which in our analysis gives $2.2\sigma$. The difference comes from the accelerator curve, since the collaboration uses a curve slightly broader in IO than the NuFIT~6.1 combination of the accelerator experiments with IceCube/DeepCore. With their curve we reproduce their result.}
and we present more details in Appendix~\ref{app:stat}.

To understand which data sets drive this preference, we can look at the combination of reactor data with each $\nu_\mu$ disappearance experiment separately.
In NO, the combined fit lands at $|\dmsqATM|=2.51\times\unit$, between the reactor and \AccIC{} best fits, which agree with each other.
In IO, the combined fit lands at $|\dmsqATM|=2.45\times\unit$, roughly in the middle of the reactor best fit at $2.48\times\unit$ and the \AccIC{} one at $2.40\times\unit$.
NOvA agrees with the reactors in NO and adds $3.6$ units to the tension in IO, disfavoring IO at $1.4\sigma$, while T2K agrees with the reactors in IO and adds $2.2$ units to the tension in NO, disfavoring NO at $0.8\sigma$.
MINOS and IceCube/DeepCore, with larger uncertainties, add about one unit of $\chi^2$ each to the tension in IO, contributing slightly to the preference for NO.
Note that \AccIC{} follows NOvA closely, with best fits at nearly the same values of $|\dmsqATM|$ in both orderings, as NOvA is the most precise of these experiments and MINOS and IceCube/DeepCore lean to the same side.

As we can see, the present preference for NO does not come from a set of consistent indications.
The two most precise accelerator experiments point to opposite orderings, and the combined result follows NOvA.
The values of $|\dmsqATM|$ preferred by T2K and NOvA differ by about $3.5\%$, roughly $1\sigma$.
This is compatible with statistical fluctuations. 
However, this difference is comparable to the $1.6$--$3.2\%$ difference between $\dmsqee$ and $\dmsqmm$ when we switch the mass ordering.

The $3.5\%$ difference between T2K and NOvA does not come from the CP phase.
The $\cos\delta_{\rm CP}$ term in Eq.~\eqref{eq:dmmm} changes $\dmsqmm$ by at most $0.4\%$, so even opposite values of $\cos\delta_{\rm CP}$ in the two experiments would account for less than $1\%$.
Appearance data may still play an indirect role.
Since the curves of T2K and NOvA come from fits to both their appearance and disappearance samples, the value of $|\dmsqATM|$ preferred by each experiment is correlated with its preferred $\theta_{23}$ and $\delta_{\rm CP}$.
The known disagreement between the two experiments in the appearance channel~\cite{Kelly:2020fkv,T2K:2025wet} may therefore partly feed into the difference in $|\dmsqATM|$, even though the difference in $\chisq$ between the orderings coming from appearance has been removed.

To understand how much a 3.5\% shift on the determination of $\dmsqATM$ by NOvA or T2K could affect the final preference for the mass ordering, we do the following exercise.
We artificially shift the T2K curve in $\dmsqATM$ to the best fit of NOvA, and vice versa. We then combine the two $\Delta\chi^2$ curves with JUNO and Daya Bay, and we estimate what would be the preference for each mass ordering, see \cref{tab:shift}.
As measured, T2K adds tension in NO and NOvA in IO, and the two partly compensate each other, leading to a mild IO rejection at $1.1\sigma$, using $CL_s$ henceforth.
With T2K's  $\Delta\chi^2$ moved to NOvA's best fit, both agree with reactors in NO, the whole tension falls on IO, which then would be rejected at $1.8\sigma$.
Conversely, with NOvA's  $\Delta\chi^2$ moved to T2K's best fit, the whole tension falls on NO, which would be rejected at $1.7\sigma$.
Note that simply summing the two $\chi^2$ curves ignores their common dependence on $\delta_{\rm CP}$ and $\theta_{23}$, so this exercise is not a combined analysis, but it shows by how much the outcome relies on the value of $|\dmsqmm|$ extracted by a single experiment.
We also call attention to the fact that the extraction of $|\dmsqmm|$ is directly correlated to the experimental capability to determine the minimum of the survival oscillation probability.
That, in turn, strongly depends on the neutrino energy reconstruction which highly depends on the physics of neutrino-nucleus interactions and the experimental treatment to assess discrepancies between event generator and near detector data~\cite{Coyle:2025xjk}.

\begin{table}[h]
\centering\small
\setlength{\tabcolsep}{4pt}
\begin{tabular}{lcccc}
\toprule
JUNO+DB+ & $\TMO$ & $\sigma_{\rm IO}$ & $\sigma_{\rm NO}$ & $\sigma_{CL_s}$ \\
\midrule
T2K + NOvA                  & $\phantom{-}1.74$ & $\phantom{-}1.3$ & $\phantom{-}0.5$ & $\phantom{-}1.1$ (IO) \\
same, at NOvA's best fit        & $\phantom{-}5.19$ & $\phantom{-}2.2$ & $-0.3$ & $\phantom{-}1.8$ (IO) \\
same, at T2K's best fit         & $-4.51$ & $-0.3$ & $\phantom{-}2.1$ & $1.7$ (NO) \\
\bottomrule
\end{tabular}
\caption{
Ordering test for JUNO+Daya Bay combined with the sum of the T2K 
and NOvA $\chisq$ curves, as measured and with one of the two curves
shifted in $\dmsqATM$ to the best fit of the other. The $\sigma$'s are
defined in Appendix~\ref{app:stat}; $\sigma_{CL_s}$ is the rejection of
the ordering in parentheses.}
\label{tab:shift}
\end{table}

\section{Future prospects}
\label{sec:future}

\begin{figure*}[t!]
\centering
\includegraphics[width=\linewidth]{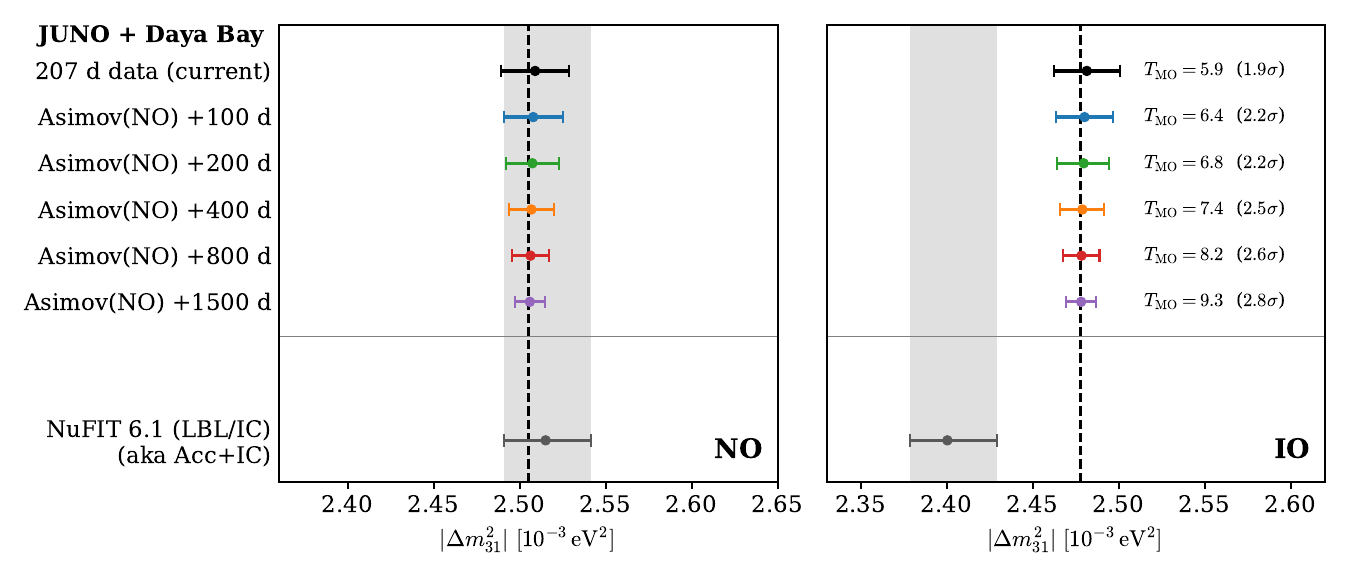}
\caption{
$1\sigma$ allowed values of $|\dmsqATM|$ in NO (left) and IO (right) 
from JUNO combined with Daya Bay, for the present 207 days of JUNO data and for
the present data combined with an Asimov spectrum of 100 to 1500 additional days
with true NO and $|\dmsqATM|=2.51\times\unit$.
The dashed lines show the limit of an exact reactor measurement, which, since the present data and systematic uncertainties remain in the fit, converges to the effective values $2.505\times\unit$ in NO and $2.478\times\unit$ in IO, and
the bottom row and the vertical gray band show the $1\sigma$ interval of the
$\nu_\mu$ disappearance data in NuFIT~6.1.
The labels in the right panel give $\TMO$ of JUNO+Daya Bay+LBL and the 
rejection of IO with $CL_s$.}
\label{fig:future_bars}
\end{figure*}

The present preference depends on two features of the JUNO data.
JUNO prefers a very low value of $|\dmsqATM|$, which produces the tension with Daya Bay found in Sec.~\ref{sec:measurements}.
Besides, due to its higher precision, JUNO determines how far the reactor value of $|\dmsqATM|$ is from the accelerator one, which generates the tension in 
IO found in Sec.~\ref{sec:combination}. 
More JUNO data will address both. 

To see how this happens, we add to the present 207 days of JUNO data an Asimov spectrum for an additional exposure $\Delta t$, computed at the true parameters, assumed to be NO with a given value of $\dmsqATM$.
We assume the signal rate to be the same as now, and we rescale the backgrounds by the extra exposure linearly. 
This is equivalent to assuming that the reactor conditions and the background rates stay the same.
This construction keeps the actual fluctuations of the present data and lets the new data pull the fit towards the true value.

We keep the measurements of Daya Bay and other experiments as they currently are, and we conservatively assume that JUNO's energy resolution and background model remain unchanged.

To start, we ask what happens if JUNO keeps preferring such a low value of the mass splitting, assuming that its true value is at the present JUNO best fit $|\dmsqATM|=1.96\times\unit$.
Further data would worsen the tension between JUNO and Daya Bay; even 200 days would already push this tension beyond the $3\sigma$ level.
Of course, we do not expect this to be realized, but this exercise shows us that, most likely, the next JUNO analysis will reveal this tension to be a statistical fluctuation.

Given that, we move to the more interesting question of how much will JUNO future data affect the mass ordering determination via the $\dmsqee$ versus $\dmsqmm$ comparison, that is, when comparing JUNO and Daya Bay data with NOvA, T2K, MINOS and IceCube/DeepCore.
Figure~\ref{fig:future_bars} shows a possible evolution of the JUNO + Daya Bay determination of $|\dmsqATM|$, assuming for true values NO and $\dmsqATM=2.51\times\unit$ (the local minimum of the JUNO $\chisq$ closest to Daya Bay's $2.51\times\unit$).

As more data significantly reduces the reactor uncertainty on the mass splitting, the preference for the mass ordering becomes more limited by the precision of the measurement of the splitting in $\nu_\mu$ disappearance.
From the current uncertainty of about $\pm0.020\times\unit$ to the possible future $\pm0.009\times\unit$ after 1500 additional days, the rejection of IO would only grow from $1.9\sigma$ to $2.8\sigma$.

In the limit of a perfect reactor measurement
(dashed line in Figure~\ref{fig:future_bars}), the rejection of IO would asymptote at $2.9\sigma$ with $CL_s$.
Therefore, more reactor data alone cannot bring the rejection of IO beyond about
$3\sigma$, since the test is then limited by the width of the accelerator
interval.
The ordering is then decided by where the accelerator interval lies with 
respect to a fixed reactor value and, as discussed in Sec.~\ref{sec:combination}, a
shift of a few percent in the value of $|\dmsqmm|$ extracted by T2K or NOvA is
enough to move the preference from one ordering to the other.

\section{Conclusions}
\label{sec:conclusion}
We have studied which data sets drive the present preference for the mass ordering from the comparison of $\dmsqee$ and $\dmsqmm$, combining JUNO and Daya Bay with $\nu_\mu$ disappearance data.
While reactor data alone show no preference, their combination with \AccIC, the NuFIT~6.1 combination of accelerator and IceCube/DeepCore data, disfavors IO at $1.9\sigma$.
This result is a balance between experiments that point to opposite orderings: when combined with reactors, NOvA disfavors IO at $1.4\sigma$, while T2K disfavors NO at $0.8\sigma$.
The values of $|\dmsqATM|$ preferred by these two experiments differ by about $3.5\%$, comparable to the difference between $|\dmsqee|$ and $|\dmsqmm|$ that distinguishes the two orderings, and therefore a shift of this size in a single experiment is enough to change the preferred ordering.
Besides, in the context of the combination of reactor and $\nu_\mu$ disappearance experiments, further JUNO data will most likely not increase the mass ordering preference beyond $3\sigma$, as the comparison is then limited by the precision of $\dmsqmm$.

The extraction of $|\dmsqmm|$ relies on the energy at which the $\nu_\mu$ survival probability reaches its minimum, and thus on the reconstruction of the neutrino energy.
Neutrino energy reconstruction at the GeV scale depends on the modeling of neutrino-nucleus interactions and on how each experiment tunes its event generator to near detector data.
As the two orderings differ by percent-level shifts in this splitting, the energy reconstruction must be understood at the same level for this comparison to determine the ordering.
The same requirement will hold for Hyper-Kamiokande~\cite{Hyper-Kamiokande:2018ofw} and DUNE~\cite{DUNE:2020ypp,DUNE:2020jqi}.
Regardless, DUNE can determine the ordering from the large impact of matter effects on the appearance probabilities.
This measurement will not rely on a percent-level comparison of mass splittings, and thus will be independent of the present hint and of combinations with other measurements.

\begin{acknowledgments}
We thank Matheus Hostert, Kevin Kelly, Shirley Li, and Stephen Parke for providing feedback on this manuscript.
J.P.P. is supported by the National Natural Science Foundation of China (12425506 and 12375101) and also by the K.C. Wong Education Foundation.
This manuscript is based upon work supported by Fermi Forward Discovery Group, LLC under Contract No. 89243024CSC000002 with the U.S. Department of Energy, Office of Science, Office of High Energy Physics. 
The work of P.M. was supported by the U.S. Department of Energy, Office of Science, Office of Advanced Scientific Computing Research, Scientific Discovery through Advanced Computing (SciDAC-5) program, grant “NeuCol”, DE-AC02-07CH11359.
J.P.P. thanks the Fermilab Theory Division for their hospitality and support through the Summer Visitors’ Program during a portion of this work. 
During this work, we used Claude Opus 5.5 (Anthropic) to assist with writing code and generating figures. 
Obviously, we take full responsibility for the results.
\end{acknowledgments}

\appendix

\section{Statistical method}
\label{app:stat}

\subsection{JUNO analysis}

Our analysis of the JUNO spectrum reproduces the one of the collaboration in Ref.~\cite{JUNO:2026tap}, based on 207 days of data.
The reactor flux, detector response, backgrounds and systematic uncertainties follow the first JUNO analyses~\cite{JUNO:2025gmd}, and our implementation is similar to the one of Ref.~\cite{Esteban:2026phq}.
We fit the spectrum in $\snqS$, $\snqR$, $\dmsqSL$ and $\dmsqATM$, and we marginalize over $\snqS$ and $\dmsqSL$, which JUNO determines from the slow oscillation.
When JUNO is shown alone, as in Fig.~\ref{fig:dm31_reactor}, we marginalize over $\snqR$ with a Gaussian prior from Daya Bay, following the collaboration.
In all combinations this prior is removed, and $\snqR$ is constrained by the two-dimensional $\dchisq$ of Daya Bay.

\subsection{Daya Bay and \AccIC{} inputs}

\begin{figure*}[t]
\centering
\includegraphics[width=0.9\linewidth]{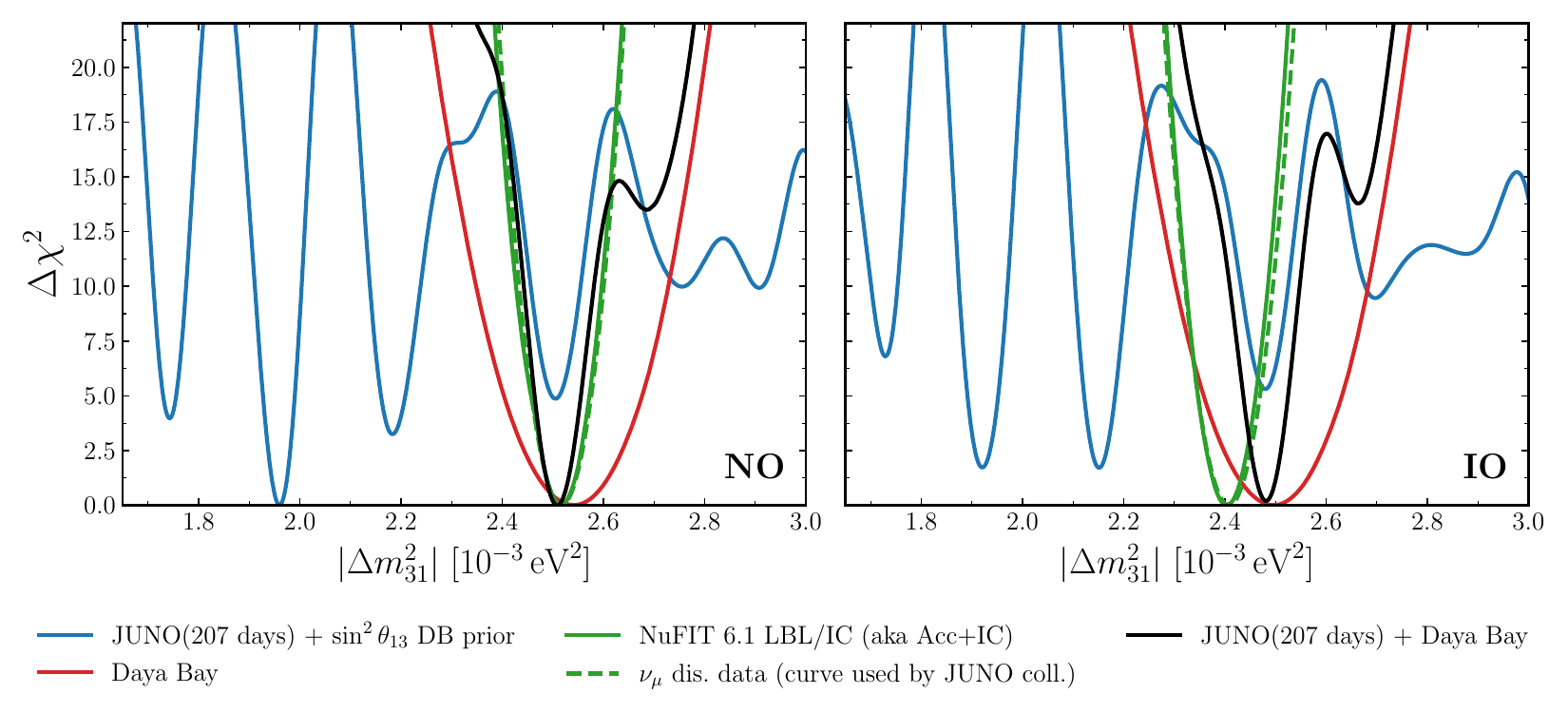}
\caption{
$\Delta\chi^2(\dmsqATM)$ in NO (left) and IO (right) for JUNO with $\snqR$ marginalized with a Gaussian prior from Daya Bay, for Daya Bay and for \AccIC.}
\label{fig:atm_mass_chisq}
\end{figure*}

Daya Bay enters through its two-dimensional $\dchisq$ in 
$(\dmsqATM,\snqR)$~\cite{DayaBay:2022orm}, and the $\nu_\mu$ data through
the one-dimensional curves of \AccIC{} and of each experiment described in
Sec.~\ref{sec:measurements}.
All of them are shifted to zero minimum in each ordering, so that $\TMO$
comes only from the comparison of the values of $|\dmsqATM|$.
Figure~\ref{fig:atm_mass_chisq} shows these inputs.
The reactor best fit moves from $|\dmsqATM|\simeq2.52\times\unit$ in NO to
$2.48\times\unit$ in IO, and the \AccIC{} one from $2.52\times\unit$ to
$2.40\times\unit$.
The two values coincide in NO and are separated by about $0.08\times\unit$ in IO.

\subsection{Tension between experiments}

For JUNO and a set of external experiments $i$ we use the parameter goodness-of-fit test~\cite{Maltoni:2003cu},
\begin{equation}
T=\min\Big[\chi^2_J+\sum_i G_i\Big]-\min\chi^2_J-\sum_i\min G_i ,
\label{eq:T}
\end{equation}
where $\chi^2_J$ is the JUNO $\chisq$ and $G_i$ is the $\dchisq$ of experiment $i$.
The last two terms are the lowest $\chisq$ that each experiment reaches on its own, and the first term is the lowest $\chisq$ they reach when they are forced to share the same $\dmsqATM$ and $\snqR$.
The difference $T$ is the tension between the experiments in units of $\chisq$, that is, the price they pay for agreeing on a common set of parameters.
We compute $T$ separately in NO and in IO.

If the $\chisq$ of every experiment were a parabola in its parameters, Wilks' theorem~\cite{Wilks:1938dza,ParticleDataGroup:2026} would fix the distribution of $T$ to a $\chisq$ distribution with $n=\sum_i P_i-P_{\rm comb}$ degrees of freedom~\cite{Maltoni:2003cu}, where $P_i$ is the number of parameters constrained by experiment $i$ (JUNO included) and $P_{\rm comb}$ is the number of parameters of the combined fit.
JUNO and Daya Bay both measure $\dmsqATM$ and $\snqR$, which gives $n=2$, \AccIC{} measures only $\dmsqATM$, so JUNO+\AccIC{} has $n=1$, and JUNO+Daya Bay+\AccIC{} has $n=3$.
For $n=1$ this reduces to the naive rule $n_\sigma=\sqrt{T}$.

The tension also sets the ordering test.
Since every external $\dchisq$ has zero minimum in each ordering, the minimum of Eq.~\eqref{eq:chisq_comb} is $\min\chisq_J({\rm O})+T_{\rm O}$, and
\begin{equation}
\TMO=\big[\min\chisq_J({\rm IO})-\min\chisq_J({\rm NO})\big]+\big(T_{\rm IO}-T_{\rm NO}\big).
\label{eq:tmo_T}
\end{equation}
The first term is the preference of JUNO alone, $1.84$, and the second is the difference between the tensions in the two orderings in Table~\ref{tab:tension}.

\subsection{Pseudo-experiments}

The JUNO $\chisq$ is not a parabola in $\dmsqATM$, and a statistical fluctuation can move its global minimum from one of its minima to another.
When this happens, $T$ becomes large even if all experiments agree on the true parameters, so neither $\sqrt{T}$ nor Wilks' theorem can be trusted.
We therefore generate pseudo-experiments in which all data sets are fluctuated around common true parameters, taken at the combined best fit in each ordering.
For JUNO, the predicted spectrum receives Poisson fluctuations in each bin, and the central values of the pull terms of the nuisance parameters are also fluctuated.
For Daya Bay and the $\nu_\mu$ data, the minimum of each $\dchisq$ is shifted by a Gaussian random amount with the width of the curve, keeping its shape.
By construction, the data sets agree on the true parameters in every pseudo-experiment, so any tension found in them is caused by fluctuations alone.
For each pseudo-experiment we compute $T$ in the same way as for the data, and $p_{\rm toys}=N(T\ge T_{\rm obs})/N_{\rm total}$ is the probability that fluctuations alone produce the observed tension.
We use 9000 pseudo-experiments for combinations of two data sets and 12000 for JUNO+Daya Bay+\AccIC.

\begin{figure*}[t]
\centering\includegraphics[width=0.9\linewidth]{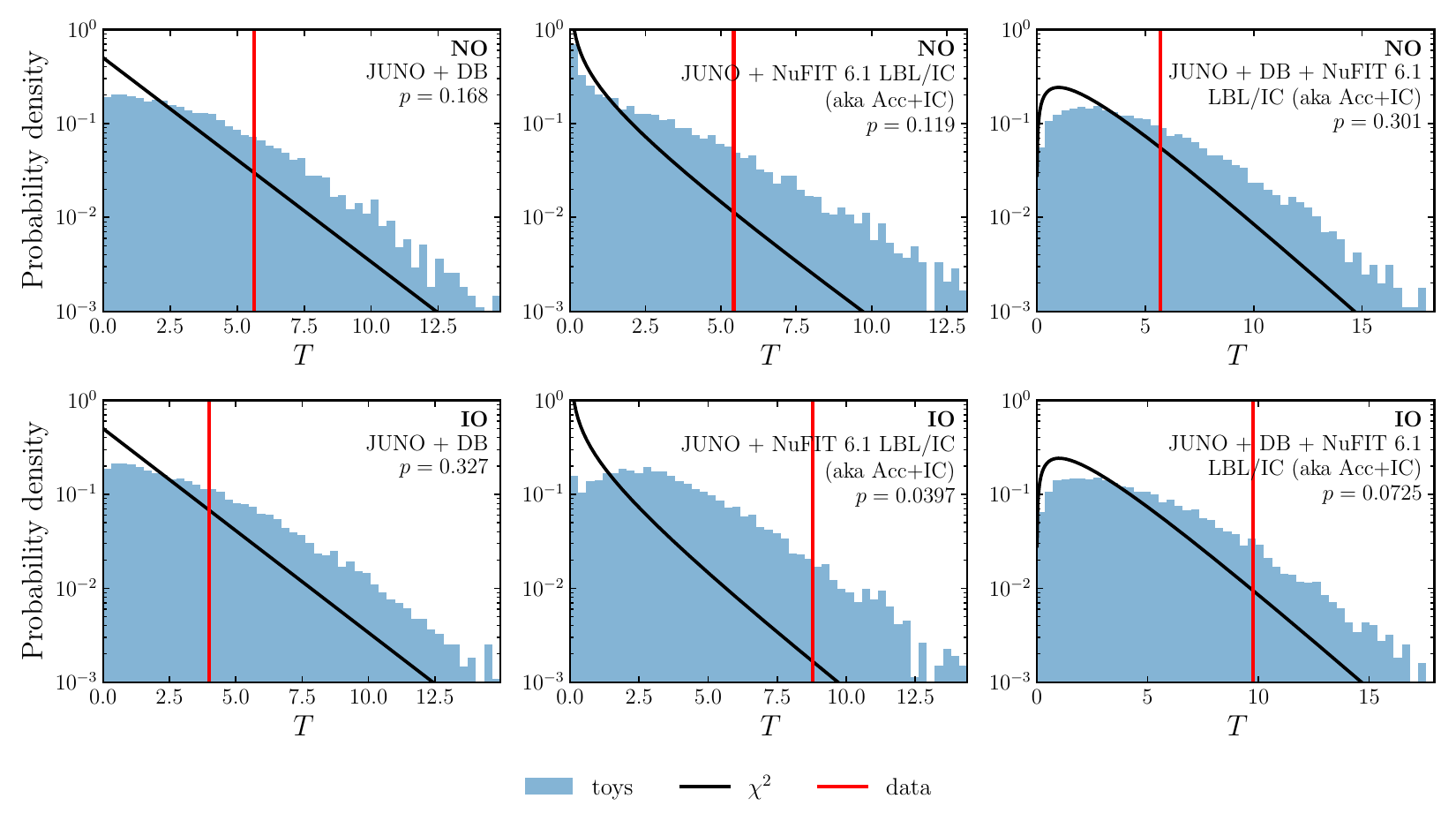}
\caption{Distribution of $T$ in pseudo-experiments for JUNO+Daya Bay+\AccIC{} in NO (left) and IO (right), compared with the $\chisq$ distribution expected from Wilks' theorem (black). The red line is the observed value.}
\label{fig:toys}
\end{figure*}

Figure~\ref{fig:toys} shows that the distributions of $T$ have a much longer tail than the $\chisq$ expectation, populated by the pseudo-experiments in which the JUNO best fit jumps to a neighbouring minimum.
Table~\ref{tab:tension} compares the three readings of the observed tension.
In NO the observed $T$ is almost the same for every combination, $T\simeq5.5$, which is the cost of moving JUNO from its global minimum to the minimum at $2.50\times\unit$.
In IO it grows from $4.0$ for JUNO+Daya Bay to $9.8$ when \AccIC{} is added, and this growth is the disagreement between $\dmsqee$ and $\dmsqmm$.
Read as $\sqrt{T}$, the tension of the full combination is $2.4\sigma$ in NO and $3.1\sigma$ in IO, while with pseudo-experiments it is $1.0\sigma$ and $1.8\sigma$.
The same effect enters the confidence intervals of JUNO, which we compute with the Feldman--Cousins construction~\cite{Feldman:1997qc}, in which the $\dchisq$ cut is obtained point by point from pseudo-experiments generated at that point instead of being fixed by Wilks' theorem.

\begin{table*}[t]\centering\small
\begin{tabular}{llccccc}
\toprule
data sets & ord. & $T$ & dof & $\sqrt{T}$ & Wilks & pseudo-exp. \\
\midrule
JUNO+Daya Bay & NO & 5.64 & 2 & 2.38$\sigma$ & 0.060 (1.9$\sigma$) & 0.168 (1.4$\sigma$) \\
JUNO+Daya Bay & IO & 4.01 & 2 & 2.00$\sigma$ & 0.135 (1.5$\sigma$) & 0.327 (1.0$\sigma$) \\
JUNO+\AccIC & NO & 5.43 & 1 & 2.33$\sigma$ & 0.020 (2.3$\sigma$) & 0.119 (1.6$\sigma$) \\
JUNO+\AccIC & IO & 8.77 & 1 & 2.96$\sigma$ & 0.0031 (3.0$\sigma$) & 0.040 (2.1$\sigma$) \\
JUNO+Daya Bay+\AccIC & NO & 5.70 & 3 & 2.39$\sigma$ & 0.127 (1.5$\sigma$) & 0.301 (1.0$\sigma$) \\
JUNO+Daya Bay+\AccIC & IO & 9.77 & 3 & 3.13$\sigma$ & 0.021 (2.3$\sigma$) & 0.072 (1.8$\sigma$) \\
\bottomrule
\end{tabular}
\caption{Tension in $\dmsqATM$ in each ordering, Eq.~\eqref{eq:T}, with the significance obtained from $\sqrt{T}$, from Wilks' theorem and from pseudo-experiments. The $p$-values are given with the corresponding two-sided significance.}
\label{tab:tension}
\end{table*}

\subsection{Mass ordering}

To decide which ordering the data prefer, we ask how often each 
ordering would produce a value of $\TMO$ like the observed one.
For this, we generate 12000 pseudo-experiments with true NO and 12000 with
true IO, in the same way as above, with the true parameters at the
combined best fit of each ordering, $|\dmsqATM|=2.510\times\unit$ in NO
and $2.450\times\unit$ in IO.
Since a large $\TMO$ favors NO, IO is disfavored when it rarely produces a
value of $\TMO$ as large as the observed one, and NO is disfavored when it
rarely produces a value as small.
These two probabilities are
\begin{align}
p_{\rm IO}&=P\big(\TMO\ge\TMO^{\rm obs}\,\big|\,{\rm IO}\big),\nonumber\\
p_{\rm NO}&=P\big(\TMO\le\TMO^{\rm obs}\,\big|\,{\rm NO}\big).
\label{eq:pmo}
\end{align}

The JUNO collaboration quotes $p_{\rm IO}$, which measures how unlikely
the data are if IO is true.
It does not check whether the data are likely if NO is true, so a small
$p_{\rm IO}$ could also come from data that are unusual for both orderings.
The $CL_s$ method~\cite{Read:2002hq,Blennow:2013oma} includes this
information by dividing the probability of the tested ordering by the
probability that the other ordering produces the observed outcome,
\begin{equation}
CL_s({\rm IO})=\frac{p_{\rm IO}}{1-p_{\rm NO}} ,\qquad CL_s({\rm NO})=\frac{p_{\rm NO}}{1-p_{\rm IO}} ,
\label{eq:cls}
\end{equation}
where $1-p_{\rm NO}$ is the probability that NO gives a $\TMO$ at least as
large as observed.
When the data are typical of NO, $1-p_{\rm NO}$ is close to one and
$CL_s({\rm IO})\simeq p_{\rm IO}$.
When the data are unusual also for NO, $1-p_{\rm NO}$ is small, 
$CL_s({\rm IO})$ grows, and the rejection of IO becomes weaker.
$CL_s({\rm NO})$ plays the same role when the data favor IO.

We convert each probability into a one-sided significance, 
$n_\sigma=\Phi^{-1}(1-p)$, with $\Phi$ the cumulative standard normal
distribution, so that $p=0.5$ corresponds to $0\sigma$ and $p=0.023$ to
$2\sigma$.
A probability larger than $0.5$ gives a negative significance.
This means that the observed $\TMO$ lies on the side of the tested
ordering, so that the data favor this ordering more than its typical
outcome does, and the ordering is not disfavored at all.
For NO, this happens when the observed $\TMO$ is above the median of the
NO distribution, and for IO, when it is below the median of the IO
distribution.

\begin{figure}[t]
\centering\includegraphics[width=\linewidth]{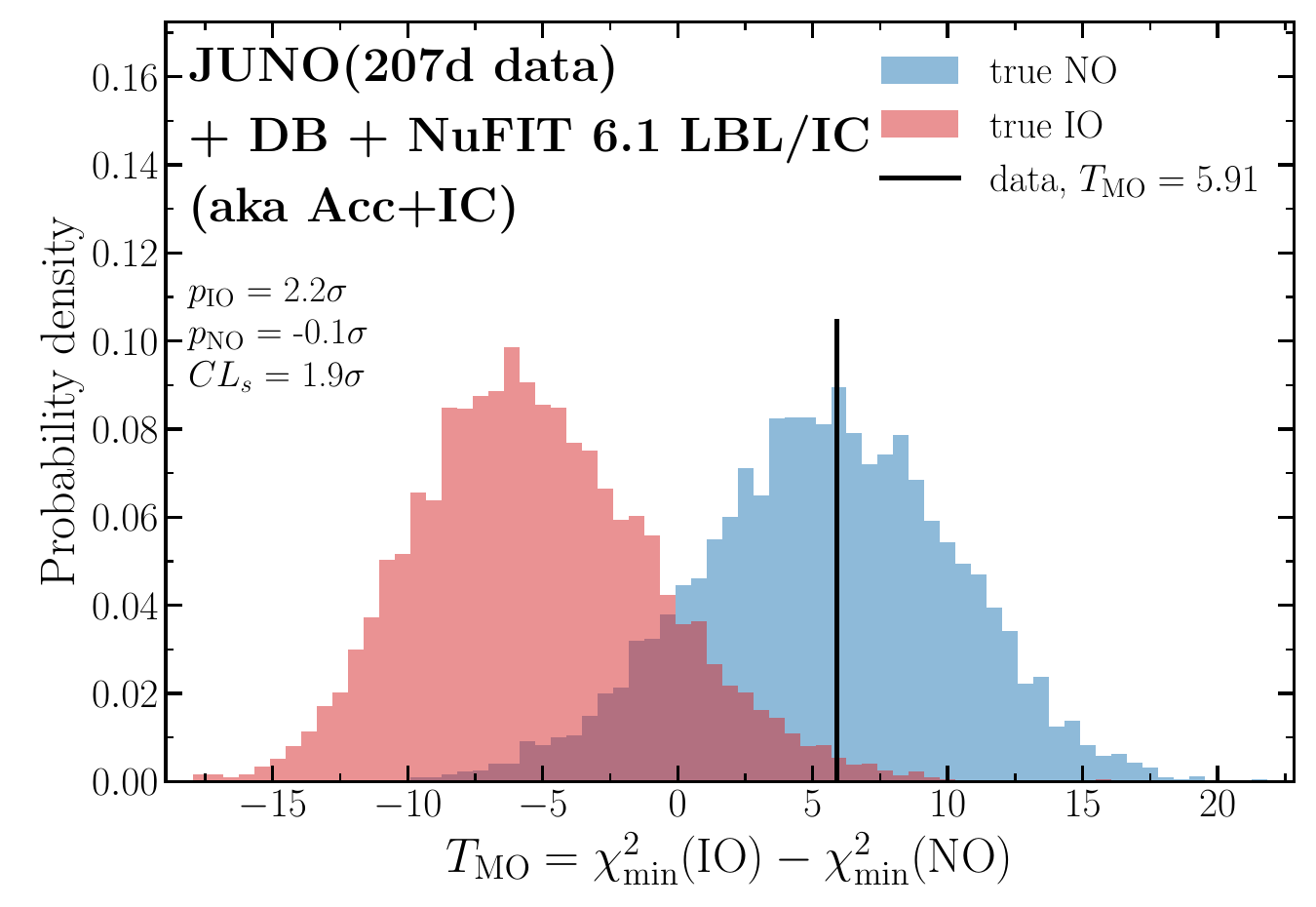}
\caption{Distribution of $\TMO$ for JUNO+Daya Bay+\AccIC{} in pseudo-experiments with true NO (blue) and true IO (red). The black line is the observed value.}
\label{fig:mo_lbl}
\end{figure}

Figure~\ref{fig:mo_lbl} shows the two distributions for JUNO+Daya Bay+\AccIC.
The observed $\TMO=5.91$ lies in the tail of the IO distribution, so that IO rarely produces it, with $p_{\rm IO}=0.013$ ($2.2\sigma$).
Since this value sits at the centre of the NO distribution, $CL_s$ only slightly weakens the rejection, with $CL_s({\rm IO})=0.027$ ($1.9\sigma$).

\subsection{Comparison with the JUNO collaboration}

The accelerator curve used by the JUNO collaboration~\cite{JUNO:2026tap} is slightly broader in IO than \AccIC.
With that curve we find $\TMO=4.85$ and $p_{\rm IO}=0.018$ ($2.1\sigma$), in agreement with the collaboration (Table~\ref{tab:official}).
With \AccIC{} the IO pseudo-experiments reach the observed value less easily, and the rejection becomes $2.2\sigma$.

\begin{table}[t]\centering\small
\setlength{\tabcolsep}{3pt}
\begin{tabular}{lccc}
\toprule
& \multicolumn{2}{c}{$\TMO$} & \\
analysis & J+DB & J+DB+acc. & $p_{\rm IO}$ \\
\midrule
JUNO collaboration~\cite{JUNO:2026tap} & 0.05 & 4.86 & 0.020 (2.1$\sigma$) \\
this work, curve of Ref.~\cite{JUNO:2026tap} & 0.21 & 4.85 & 0.018 (2.1$\sigma$) \\
this work, \AccIC & 0.21 & 5.91 & 0.013 (2.2$\sigma$) \\
\bottomrule
\end{tabular}
\caption{Comparison with the JUNO collaboration. The column J+DB+acc.\ combines JUNO and Daya Bay with the accelerator curve of each row.}
\label{tab:official}
\end{table}

\bibliographystyle{apsrev4-2}
\bibliography{references}

\end{document}